\documentclass[%
 aip,
 amsmath,amssymb,
 reprint,%
]{revtex4-1}

\usepackage{graphicx}% Include figure files
\usepackage{dcolumn}% Align table columns on decimal point
\usepackage{bm}% bold math
\usepackage{gensymb}
\usepackage{color}
\begin{document}

%\preprint{AIP/123-QED}

\title[{\color{blue}\textbf{Published in App. Phys. Lett. 112, 031109 (2018). https://doi.org/10.1063/1.5017108}}]{Polarized micro-Raman studies of femtosecond laser written stress-induced optical waveguides in diamond}% Force line breaks with \\
%\thanks{Footnote to title of article.}

\author{B. Sotillo}
\email{bsotillo@gmail.com}
\affiliation{Istituto di Fotonica e Nanotecnologie-Consiglio Nazionale delle Ricerche (IFN-CNR) and Department of Physics, Politecnico di Milano, Piazza Leonardo da Vinci 32, 20133, Milano, Italy}%
\author{A. Chiappini}
 \affiliation{Istituto di Fotonica e Nanotecnologie-Consiglio Nazionale delle Ricerche (IFN-CNR) Characterization and Development of Materials for Photonics and Optoelectronics (CSMFO) and The Centre for Materials and Microsystems (FBK-CMM) 38123 Trento Italy}%
\author{V. Bharadwaj}
\affiliation{Istituto di Fotonica e Nanotecnologie-Consiglio Nazionale delle Ricerche (IFN-CNR) and Department of Physics, Politecnico di Milano, Piazza Leonardo da Vinci 32, 20133, Milano, Italy}
\author{J.P. Hadden}
\affiliation{Institute for Quantum Science and Technology, University of Calgary, AB T2N 1N4 Calgary Canada}
\author{F. Bosia}
\affiliation{Department of Physics and “Nanostructured Interfaces and Surfaces” Inter-Departmental Centre, University of Torino, I-10125 Torino Italy}
\author{P. Olivero}
\affiliation{Department of Physics and “Nanostructured Interfaces and Surfaces” Inter-Departmental Centre, University of Torino, I-10125 Torino Italy}
\author{M. Ferrari}
\affiliation{Istituto di Fotonica e Nanotecnologie-Consiglio Nazionale delle Ricerche (IFN-CNR) Characterization and Development of Materials for Photonics and Optoelectronics (CSMFO) and The Centre for Materials and Microsystems (FBK-CMM) 38123 Trento Italy}
\author{R. Ramponi}
\affiliation{Istituto di Fotonica e Nanotecnologie-Consiglio Nazionale delle Ricerche (IFN-CNR) and Department of Physics, Politecnico di Milano, Piazza Leonardo da Vinci 32, 20133, Milano, Italy}
\author{P.E. Barclay}
\affiliation{Institute for Quantum Science and Technology, University of Calgary, AB T2N 1N4 Calgary Canada}
\author{S.M. Eaton}
\affiliation{Istituto di Fotonica e Nanotecnologie-Consiglio Nazionale delle Ricerche (IFN-CNR) and Department of Physics, Politecnico di Milano, Piazza Leonardo da Vinci 32, 20133, Milano, Italy}


\begin{abstract}
Understanding the physical mechanisms of the refractive index modulation induced by femtosecond laser writing is crucial for tailoring the properties of the resulting optical waveguides. In this work we apply polarized Raman spectroscopy to study the origin of stress-induced waveguides in diamond, produced by femtosecond laser writing. The change in the refractive index induced by the femtosecond laser in the crystal is derived from the measured stress in the waveguides. The results help to explain the waveguide polarization sensitive guiding mechanism, as well as providing a technique for their optimization. 
%Valid PACS numbers may be entered using the \verb+\pacs{#1}+ command.
\end{abstract}

\maketitle

%\begin{quotation}
%\verb+quotation+ environment and is formatted as a single paragraph before the first section heading. 
%(The \verb+quotation+ environment reverts to its usual meaning after the first sectioning command.) 
%Note that numbered references are allowed in the lead paragraph.
%
%The lead paragraph will only be found in an article being prepared for the journal \textit{Chaos}.
%\end{quotation}

Color centers in diamond, such as nitrogen-vacancy (NV) or silicon-vacancy (SiV) centers, show great potential for quantum systems \cite{aharonovich2011diamond}, temperature sensing\cite{neumann2013high,nguyen2017all} or magnetic field sensing in the case of the NV center \cite{schirhagl2014nitrogen}. These defects can be initialized, manipulated and read out using photons. Optical waveguides in bulk diamond could be used to optically link and address these color centers \cite{aharonovich2011diamond2}. Several methods have been used for fabricating waveguides in diamond, such as ion beam assisted lift-off \cite{olivero2005ion}, plasma etching \cite{khanaliloo2015single,burek2012free} or ion implantation \cite{lagomarsino2010evidence}. Recently, optical waveguides \cite{sotillo2016diamond,courvoisier2016inscription} and Bragg gratings \cite{Bharadwaj2017Bragg} in bulk diamond have been formed using femtosecond laser technique, opening the possibility of creating 3D photonic circuits in this material. A deeper understanding of the physics underlying the writing of optical waveguides in diamond will help in the development of advanced devices integrating photonics circuits and color centers \cite{hadden2017waveguide}.
 
Femtosecond laser writing relies on the nonlinear absorption of focused ultrashort pulses, which leads to a localized modification in the bulk of transparent materials \cite{davis1996writing,eaton2005heat,gattass2008femtosecond}. In crystals such as diamond, the laser interaction typically produces a decrease in the refractive index due to the damage of the lattice, so the strategy for fabricating the waveguides is to write two lines separated by several microns and confine the optical mode between these barriers \cite{chen2014optical,sotillo2016diamond}. In such waveguides, the increase in the refractive index is associated with the stress induced between the two lines due to the volume variation of the material inside the laser-written modifications \cite{burghoff2007origins}. 

Micro-Raman spectroscopy has been previously applied to study the waveguides formed in different crystals \cite{eaton2008raman,rodenas2009anisotropic,benayas2011ultrafast}.  However to date, no works have managed to describe the relation between the polarized Raman signal, the stress induced in the waveguides and its relation with the change in the refractive index. Refracted near field profilometry is the only non-destructive measurement technique for direct quantitative characterization of the cross sectional refractive index profile of bulk waveguides \cite{oberson1998refracted,eaton2011high}. However, this technique requires index matching oil (maximum refractive index 1.8), incompatible with the refractive index of diamond (2.4). Although the quantitative phase microscopy method would allow the measurement of the refractive index change in waveguides within high refractive index materials \cite{ferraro2007quantitative}, its complexity and destructive nature make it undesirable for the characterization of diamond. On the other hand, confocal polarized $\mu$Raman spectroscopy is a non destructive technique with micrometer spatial resolution and is sensitive to the local stresses present in the material, and thus will greatly benefit the understanding of the stress-induced waveguides in diamond, as well as in other crystalline materials.

The diamond samples used in this work were synthetic CVD grown single crystal diamond, type IIa, with a dimension of 5 $\times$ 5 $\times$ 0.5 mm$^{3}$. They were purchased from MB optics, with all the facets polished and with an orientation of 4pt (\{100\}-planes) for the top and bottom larger surfaces and 2pt (\{110\}-planes) for the side surfaces. If we define the crystal axis system as X = [100], Y = [010] and Z = [001] (Fig. 1(a)), the coordinate system for the sample will be defined as X' = [110], Y' = [$\bar{1}$10], Z' = [001], as shown in Fig. 1(b), being the X'Y'Z' obtained by a 45$\degree$ rotation of the XYZ coordinates around Z axis.

The femtosecond laser used for writing optical waveguides in diamond was a regeneratively amplified Yb:KGW system (Pharos, Light Conversion) with 230-fs pulse duration, 515-nm wavelength, focused with a 1.25-NA oil immersion lens (RMS100X-O 100$\times$ Olympus Plan Achromat Oil Immersion Objective). The repetition rate of the laser was 500 kHz and the pulse energy was 100 nJ. Computer-controlled, 3-axis motion stages (ABL-1000, Aerotech) were used to translate the sample relative to the laser with a scan speed of 0.5 mm/s. In the X'Y'Z' coordinate system shown in Figure 1b, the laser is incident along the [00$\bar{1}$] direction and the modification lines are formed along the [110] direction, and the linear polarization of the laser is along the [$\bar{1}$10] direction. Overhead and transversal microscope images (obtained with a Nikon Eclipse ME600 microscope) of the laser-written lines are shown in Fig. 1(c), indicating also the X'Y'Z' coordinate system. To measure the near-field waveguide optical mode profile, light was coupled to the waveguides using a single-mode fiber and the output was imaged with a 60$\times$ aspheric lens (5721-H-B, Newport) on a beam profiler (SP620U, Spiricon). A mode field diameter of 10 $\mu$m was measured for a wavelength of 635 nm for lines separated by 13 $\mu$m. Further details of the waveguide characterization can be found in Ref.[10].

\begin{figure} [h]
	\includegraphics[width=0.45\textwidth]{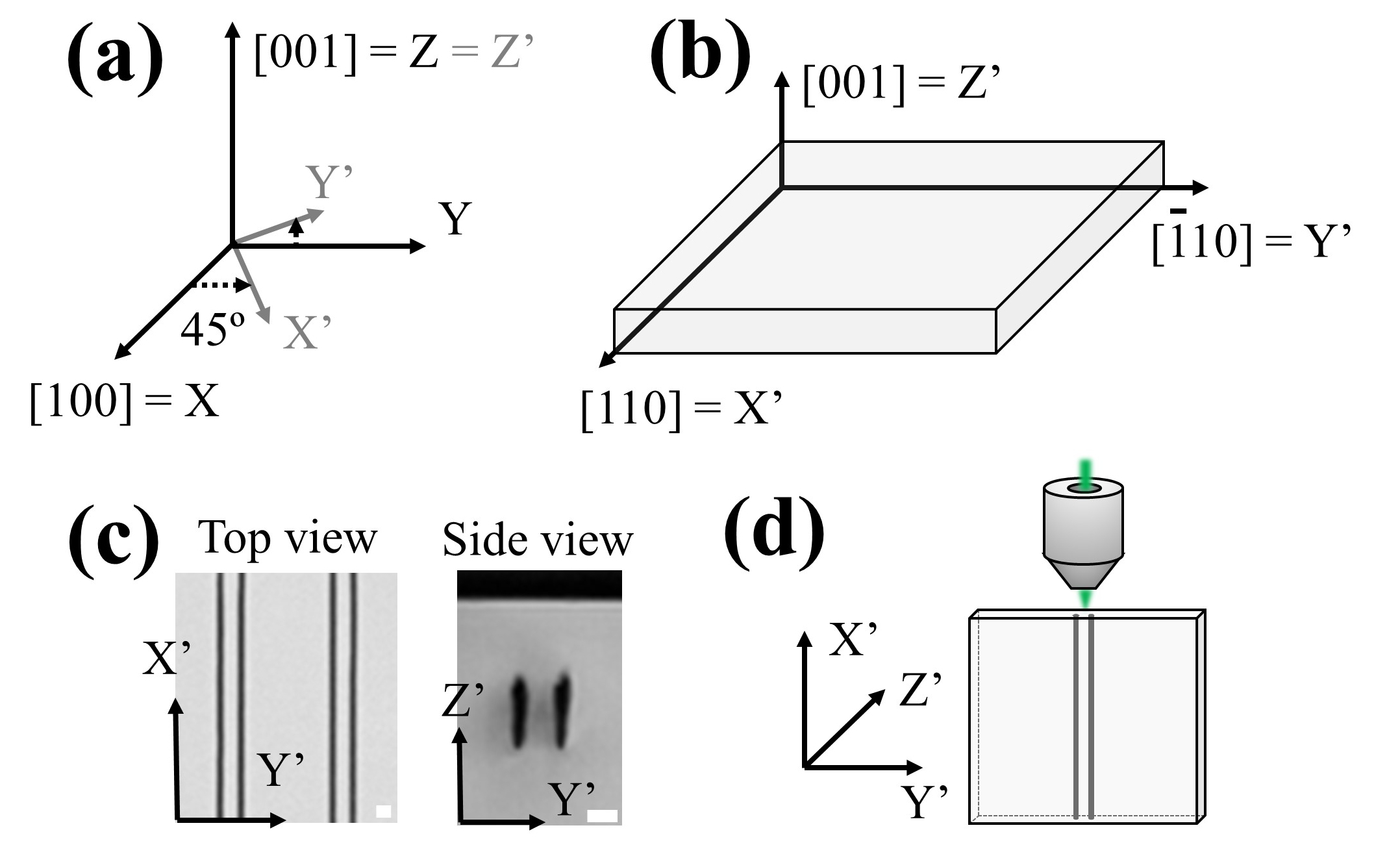}% Here is how to import EPS art
	\caption{\label{fig:epsart} Orientation of the diamond samples examined. (a) crystal axis system; (b) sample axis system. (c) Optical micrographs of the diamond waveguide (scale bar of 10 $\mu$m). (d) Sample orientation in the Raman setup.}
\end{figure}

Finally, $\mu$Raman spectra were recorded using a Labram Aramis Jobin Yvon Horiba microRaman system with a DPSS laser source of 532 nm and equipped with a confocal microscope and an air-cooled CCD. The pinhole was 300 $\mu$m. A 100$\times$ objective (0.8 NA) was used to focus the laser on the (110) plane of the sample as well as to collect the Raman signal (backscattering configuration), as shown in Fig. 1(d). This experimental configuration allows to detect the Raman modes needed to perform this study. The spatial resolution is below 1 $\mu$m.

Cubic diamond lattice (space group $ O_{h} $) has three optical phonons with $ F_{2g} $ symmetry at the center of the Brillouin zone. These phonons, in absence of strain in the lattice, have the same frequency ($\omega_{0}$ = 1333 cm$^{-1}$). The Raman intensity collected is given by the following expression, that depends on the Raman tensors ($R_{j}$) and the polarization vectors of the incident ($ e_{i} $) and scattered ($ e_{s} $) light:
\begin{equation}
I \propto \sum |e_{i} \cdot R_{j} \cdot e_{s}|^{2}\label{RamanI}
\end{equation} 
The Raman tensors at the center of the Brillouin zone in the XYZ crystal system are: 
\[
R_{1}=
\begin{pmatrix}
0 & 0 & 0 \\
0 & 0 & d \\
0 & d & 0 \\
\end{pmatrix};
R_{2}=
\begin{pmatrix}
0 & 0 & d \\
0 & 0 & 0 \\
d & 0 & 0 \\
\end{pmatrix};
R_{3}=
\begin{pmatrix}
0 & d & 0 \\
d & 0 & 0 \\
0 & 0 & 0 \\
\end{pmatrix}
\]

Where $d$ accounts for the variation of the polarizability with the displacement of the atoms in the lattice \cite{mildren2013optical}. For the coordinate system of our diamond sample X'Y'Z' these tensors will be transformed as (rotation of 45$\degree$ around the Z axis):
\[
R_{1}'=d
\begin{pmatrix}
0 & 0 & -1/\sqrt{2} \\
0 & 0 & 1/\sqrt{2} \\
-1/\sqrt{2} & 1/\sqrt{2} & 0 \\
\end{pmatrix};
\]
\[
R_{2}'=d
\begin{pmatrix}
0 & 0 & 1/\sqrt{2} \\
0 & 0 & 1/\sqrt{2} \\
1/\sqrt{2} & 1/\sqrt{2} & 0 \\
\end{pmatrix};
R_{3}'=d
\begin{pmatrix}
-1 & 0 & 0 \\
0 & 1 & 0 \\
0 & 0 & 0 \\
\end{pmatrix}
\]

It is well known that when stress is applied to the diamond crystal, the cubic symmetry is removed and the degeneracy of the three optical phonons is lifted \cite{ganesan1970lattice,grimsditch1978effect,ager1993quantitative,von1997stress,anastassakis1999strain,gries2007stresses}. In order to obtain the new frequencies of the phonons at the zone-center related to the applied stress, the following secular equation, that derives from lattice dynamical equations \cite{ganesan1970lattice} has to be solved: 

\begin{widetext}
	\begin{equation}
	\begin{vmatrix}
	p\epsilon_{1}+q(\epsilon_{2}+\epsilon_{3})-\lambda & r\epsilon_{6} & r\epsilon_{5}\\
	r\epsilon_{6} & p\epsilon_{2}+q(\epsilon_{1}+\epsilon_{3})-\lambda & r\epsilon_{4}\\
	r\epsilon_{5} & 2r\epsilon_{4} & p\epsilon_{3}+q(\epsilon_{1}+\epsilon_{2})-\lambda\\\
	\end{vmatrix}=0 
	\label{secular}
	\end{equation} 
\end{widetext}

Where $p$, $q$ and $r$ are the deformation potential constants for diamond \cite{grimsditch1978effect,gries2007stresses} ($p=-2.810\omega_{0}^{2};~ q=-1.77\omega_{0}^{2};~ r=-1.990\omega_{0}^{2}$), and the eigenvalues $\lambda_{i}$ are related to the former ($\omega_{0}$) and new frequencies of the three modes ($\omega_{i}$) \cite{von1997stress}: $\lambda_{i}=\omega_{i}^{2}-\omega_{0}^{2};
~\Delta\omega_{i}=\omega_{i}-\omega_{0}\approx\dfrac{\lambda_{i}}{2\omega_{0}}$.
$\epsilon$ represents the components of the strain matrix, that are related to the stress $\sigma$ by Hooke's law: $\epsilon_{i}=S_{ij}\sigma_{j}$ (i,j = 1,..., 6, being $S_{ij}$ the compliance constants, following the matrix notation \cite{nye1985physical}).

For a cubic crystal, due to symmetry considerations, only three components are non-zero, having the following values for diamond \cite{gries2007stresses}: $S_{11}= 0.952\times10^{-3}~$GPa;~ $S_{12}= -0.099\times10^{-3}$~GPa;~ $S_{44}= 1.737\times10^{-3}~$GPa. As noted above, in the selected scenario of our waveguide, we are going to study the Raman signal in backscattering configuration from the (110) surface, considering that the stress produced by the two laser-written tracks has biaxial character in the (110) plane. The stress matrix referred to the X'Y'Z' system is defined then as:
\begin{equation}
\sigma'=
\begin{pmatrix}
0 & 0 & 0 \\
0 & \tau_{1} & 0 \\
0 & 0 & \tau_{2} \\
\end{pmatrix}
\end{equation}
Where $\tau_{1}$ is the stress parallel to the [$\bar{1}$10] direction and $\tau_{2}$ is the stress parallel to the [001] direction. This matrix can be rotated into the XYZ crystal coordinates by applying a rotation of -45$\degree$ around Z' = Z axis:
\begin{equation}
\sigma=
\begin{pmatrix}
\tau_{1}/{2} & \tau_{1}/{2} & 0 \\
\tau_{1}/{2} & \tau_{1}/{2} & 0 \\
0 & 0 & \tau_{2} \\
\end{pmatrix}
\end{equation}
Using this matrix, the strain components are calculated using Hooke's law:
$
\epsilon_{1}=\epsilon_{2}=s_{12}\tau_{2}+\dfrac{\tau_{1}}{2}(s_{11}+s_{12});
~\epsilon_{3}=s_{11}\tau_{2}+s_{12}\cdot\tau_{1};
~\epsilon_{6}=\dfrac{s_{44}\tau_{1}}{2}.
$ By substituting in the secular equation (\ref{secular}), the eigenvalues are obtained:
\begin{equation}
\Delta\omega_{1}=-(0.82\tau_{2}+0.085\tau_{1}), \label{shift1}
\end{equation}
\begin{equation}
\Delta\omega_{2}=-(0.82\tau_{2}+2.28\tau_{1}), \label{shift2}
\end{equation}
\begin{equation}
\Delta\omega_{3}=-(1.55\tau_{2}+0.82\tau_{1}),\label{shift3}
\end{equation}
In the case of $\tau_{1}=\tau_{2}$ the values reported by Von Kaenel et al. \cite{von1997stress} are reproduced. These results indicate that the applied stress lifted completely the triple degeneracy. Note that the sign convention for stress is that tensile stresses are positive while compressive are negative. 
From equation (\ref{RamanI}) and considering the conservation of the wavevector, we can derive the components that will be visible for backscattering from a (110) surface, which are shown in table I, in agreement with other studies \cite{anastassakis1970effect,von1997stress}  (note that Porto's notation is used \cite{arguello1969first}).

\begin{table}[h]
	\caption{\label{tab:table1}Selection rules - Backscattering from (110)}
	\begin{ruledtabular}
		\begin{tabular}{lcr}
			Geometry&Component //[001]&Component //[$\bar{1}$10]\\
			\hline\\
			$\bar{X}$'(Y'Y')X' & \checkmark & -\\
			$\bar{X}$'(Y'Z')X' & - & \checkmark\\
			$\bar{X}$'(Z'Y')X' & - & \checkmark\\
			$\bar{X}$'(Z'Z')X' & - & -\\
		\end{tabular}
	\end{ruledtabular}
\end{table}

From previous studies \cite{ager1993quantitative}, the out-of-plane component will show the most different value of shift. So the values obtained in equation (\ref{shift1}) will be related to this component, whereas equation (\ref{shift2}) and (\ref{shift3}) are the components that will be measured in our experiments (//[001] and //[$\bar{1}$10] components).

In Fig. 2 the maps for the intensity and the shift of the diamond peak (from the fitting to Lorenztian curves) are shown. The Raman peak shows an intensity decrease in the laser-written tracks associated with disorder and change to the sp$^{2}$-like phase induced by the laser \cite{sotillo2016diamond}. The results obtained within the modified laser regions are not physically meaningful in this study as the fit of the diamond peak is more complex due to the presence of disorder and non-diamond carbon phases, so the lines position is marked in black in the maps of Fig. 2(b,c) by overlapping the region where the peak intensity is below 20\%. In the region between the two lines, a small homogeneous broadening of the Raman peak from a width of 2.1 cm$^{-1}$ to a width of 2.4 cm$^{-1}$ is detected, related to the presence of stress \cite{bergman1995raman,erasmus2011application}. The linewidth is similar along the guiding region, which is an indication of fairly uniform stress distribution.

\begin{figure} [h]
	\includegraphics[width=0.45\textwidth]{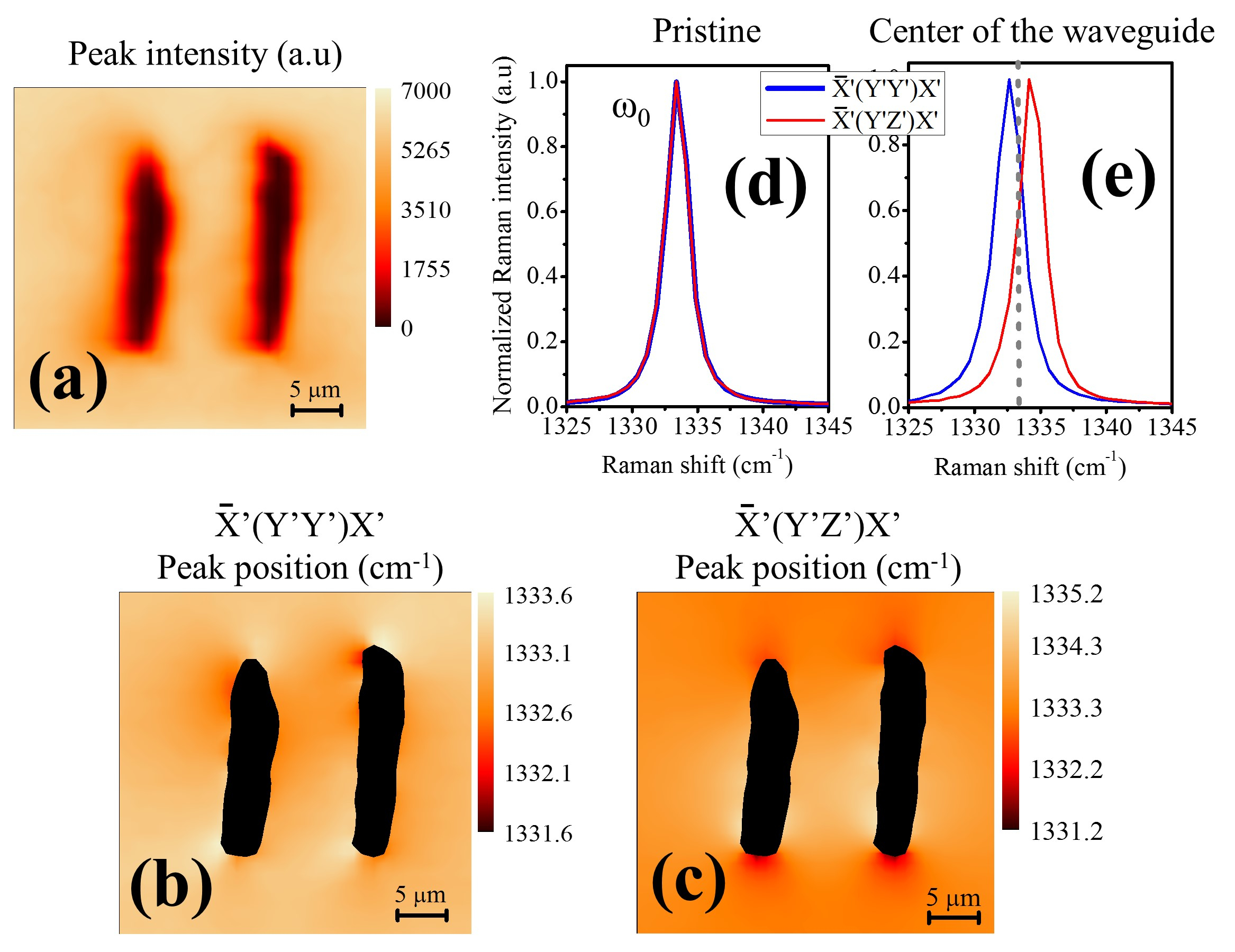}% Here is how to import EPS art
	\caption{\label{fig:epsart} Maps of the (a) intensity and the shift of the diamond Raman peak in  the (b) $\bar{X}$'(Y'Y')X' and (c) $\bar{X}$'(Y'Z')X' configurations, obtained from the fit of the spectra to Lorentzian curves (scanning step of 1 $\mu$m). Raman spectra recorded (d) outside and (e) within waveguide region.%, where the shift to lower/higher wavenumber visible.
	}
\end{figure}

Following the selection rules of table I, if the configuration $\bar{X}$'(Y'Y')X' is selected, the component parallel to [001] can be obtained, and is the one presented in Fig. 2(b). On the other hand, $\bar{X}$'(Y'Z')X' configuration is shown in Fig. 2(c), which is associated with the component parallel to [$\bar{1}$10]. The shift of these two components (Fig. 2(d,e)) is a first indication that we can have stress with different character (tensile or compressive) in the [001] and [$\bar{1}$10] directions. In order to calculate the stress responsible for these shifts, equations (\ref{shift2}) and (\ref{shift3}) can be applied. A equation system must be solved in order to extract the values of $\tau_{1}$ and $\tau_{2}$ from the measured shifts:
\begin{equation}
\tau_{1}=-\dfrac{\Delta\omega_{2}-0.53\Delta\omega_{3}}{1.85};
~\tau_{2}=-\dfrac{\Delta\omega_{3}+0.82\tau_{1}}{1.55}
\end{equation}

At this point we have to consider the effect that the two laser lines can have on the surrounding material. The laser causes the amorphization of the crystal lattice and the transformation of sp$^{3}$-diamond bonding into sp$^{2}$ bondings \cite{sotillo2016diamond}. This transformation generates a material with lower density inside the lines, producing a local volume expansion. Furthermore, the stress produced in the [$\bar{1}$10] direction ($\tau_{1}$) will be compressive, whereas in the perpendicular direction there will be a shear tensile stress ($\tau_{2}$). Taking this into account, we associate the correct component $\Delta\omega_{2}$ and $\Delta\omega_{3}$ to the measured Raman components, and the calculated stress for each point in the map is shown in Fig. 3(a,b). In these maps it can be observed that $\tau_{1}$ has a compressive character between the two lines, whereas $\tau_{2}$ is mainly tensile. The sign of both stresses is inverted at the top and bottom tips of the lines (in a similar way to the theoretical study covering lithium niobate  \cite{burghoff2007origins}). From the vertical line profile of both stresses $\tau_{1}$ and $\tau_{2}$ measured at the center of the waveguide (Fig. 3(c)), it can be observed that the maximum value for both is located near the center of the waveguide.

\begin{figure} [h]
	\includegraphics[width=0.48\textwidth]{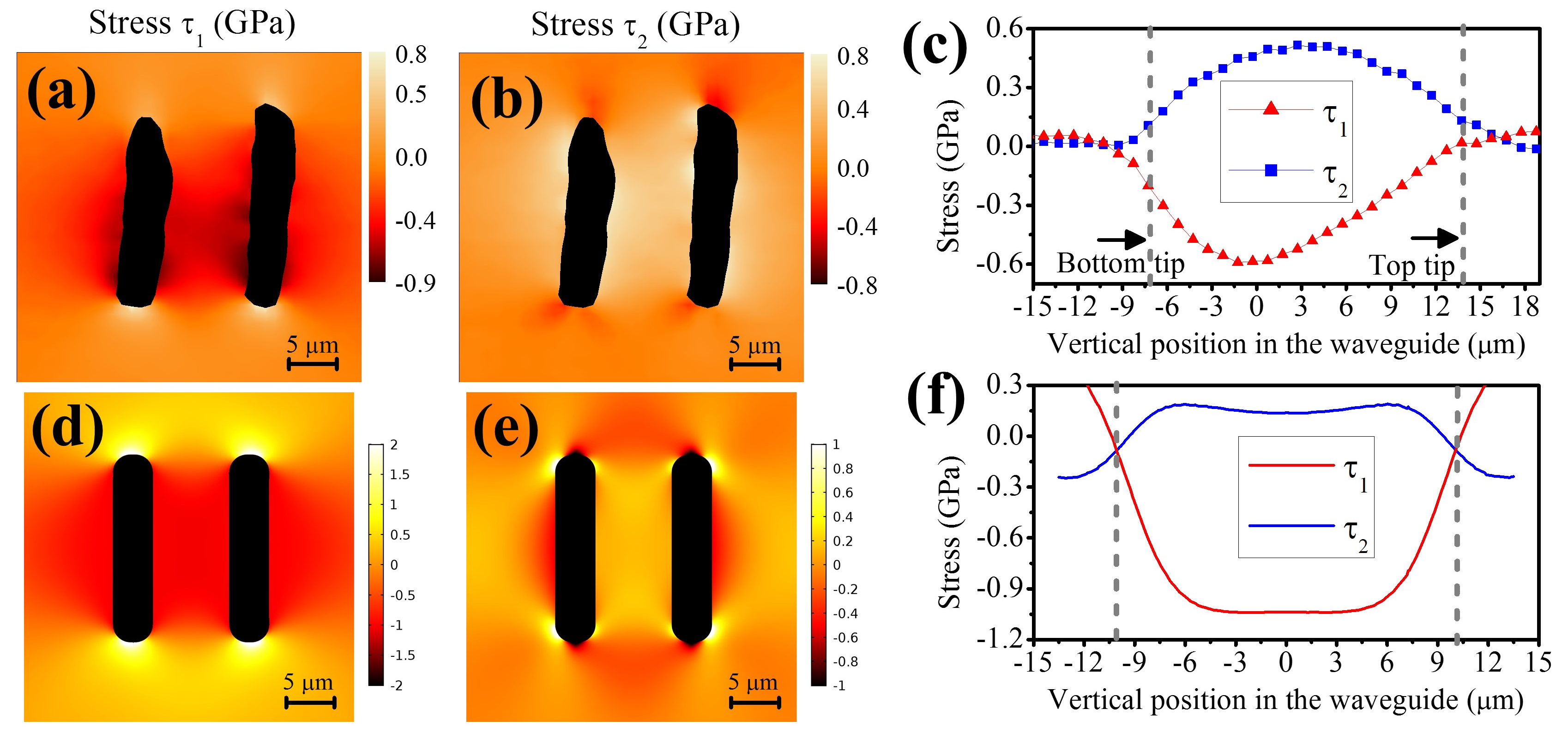}% Here is how to import EPS art
	\caption{\label{fig:epsart} Maps of the $\tau_{1}$ and $\tau_{2}$ induced stresses in the waveguide: (a)-(b) from Raman peak shift and (d)-(e) form 2D Finite Elements simulations. Vertical line profile of $\tau_{1}$ and $\tau_{2}$ at the center of the waveguide: (c) measured and (f) simulated.}
\end{figure}

The calculated stresses are compared to those obtained from 2D Finite Element simulations of the constrained expansion of two amorphous carbon inclusions in a diamond matrix. The adopted procedure is described in Bosia \textit{et al.}\cite{bosia2013direct} and in Battiato \textit{et al.}  \cite{battiato2016softening}. To approximate the real geometry, two parallel rectangles with filleted corners are considered. The simulated $\tau_{1}$ and $\tau_{2}$ stress distributions, shown in Fig. 3 (d-f), correctly reproduce those observed experimentally. The small discrepancies and the asymmetry between the curves of $\tau_{1}$ and $\tau_{2}$ in Fig. 3(c) can be attributed to the simplified geometry in the simulations and possible inhomogeneities in material properties in the amorphized regions.

Once the stresses have been calculated from the measured shift of the Raman peak, the change in the refractive index can be obtained following the same strategy as Burghoff \textit{et al.} \cite{burghoff2007origins}. Stress values are related to the refractive index through the piezooptic tensor $\pi_{ij}$:
\begin{equation}
\Delta\left(\dfrac{1}{n^{2}}\right)_{i}=\pi_{ij}\sigma_{j} \label{piezooptic}
\end{equation}
As for the compliance tensor, due to the symmetry of the diamond crystal structure, only three components of the piezooptic tensor are non-zero: $\pi_{11}= -0.43\times10^{-12}~$Pa;~ $\pi_{12}= 0.37\times10^{-12}$~Pa;~ $\pi_{44}= -0.27\times10^{-12}~$Pa \cite{nye1985physical}.
Previous experimental studies have shown that tensile hydrostatic pressure produces an increase in the refractive index of diamond \cite{waxler1965effect,fontanella1977temperature,balzaretti1996pressure}. This effect is related to an increment of the electronic polarizability overbalancing the reduction of the density \cite{waxler1965effect}, and has been observed in other materials, such as MgO, SiC or sapphire \cite{waxler1965effect,balzaretti1996pressure,jones2001refractive}. Here we have a biaxial stress, so the different components of the $\Delta(1/n^{2})_{i}$ tensor must be calculated to assess the character of the refractive index change. Using equation (\ref{piezooptic}) and the values of stress $\tau_{1}$ and $\tau_{2}$, we can obtain the refractive index profile across the waveguide for the polarization parallel to [001] (Fig. 4(a)). A well localized region of increased refractive index is visible. Comparing the map of the refractive index change with the image of an optical mode (Fig. 4(c)), we can see that the mode is located at the same position as the increased refractive index region. For the polarization parallel to [$\bar{1}$10] the refractive index decreases in the guiding region.

From the map in Fig. 4(a), we can see that the optical mode is confined in the horizontal direction by the presence of the two laser-written lines which act as barriers due the decrease $n$ expected in these regions. In vertical direction, light is confined due to the stress-induced change in the refractive index. At the center of the waveguide, the $\Delta n$ calculated is $3\times10^{-3}$. 

\begin{figure} [h]
	\includegraphics[width=0.35\textwidth]{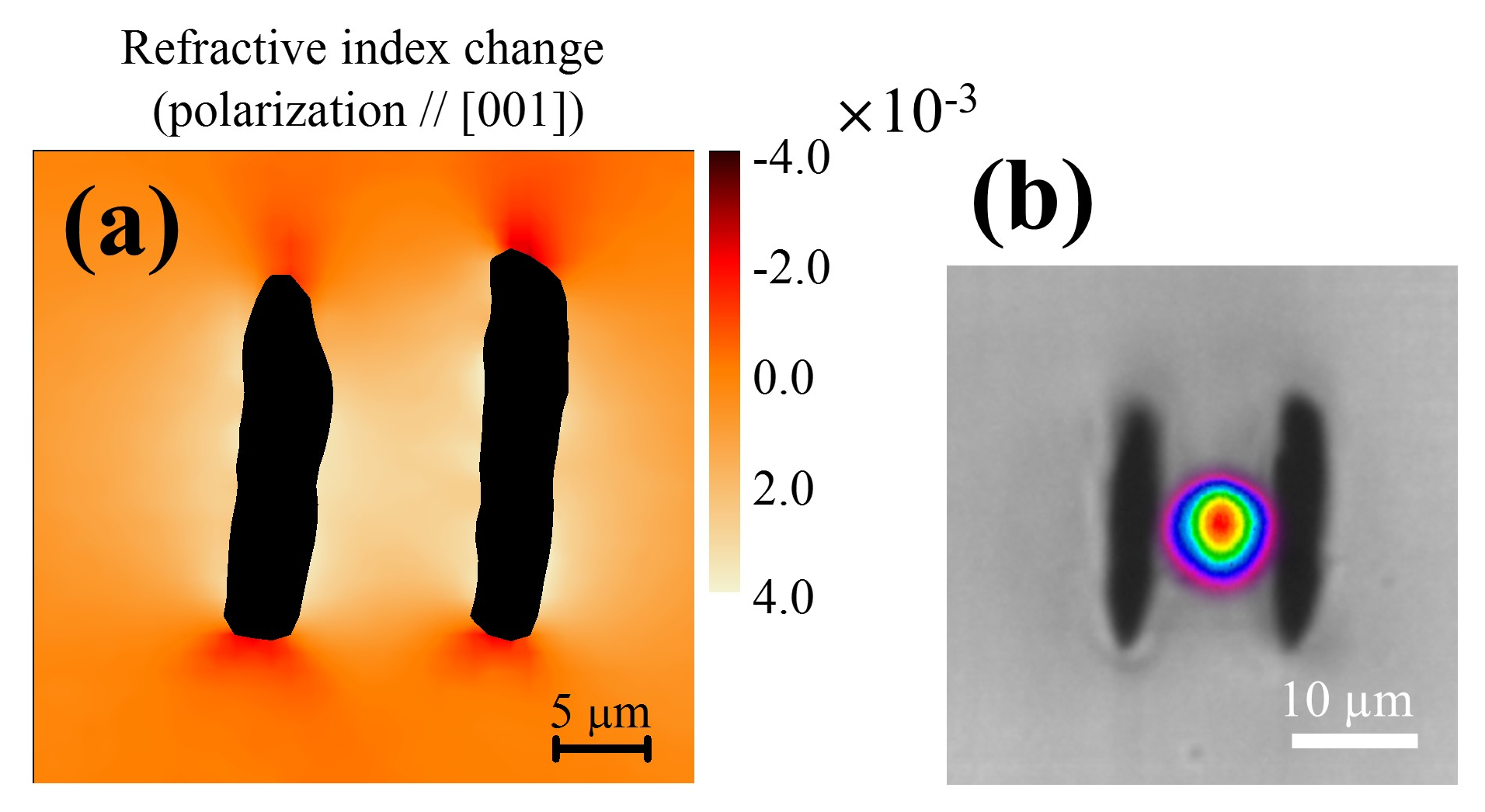}% Here is how to import EPS art
	\caption{\label{fig:epsart} (a) Map of the refractive index profile. (b)  Optical microscope image with overlaid optical TM mode measured with a beam-profiler (mode field diameter of $\sim$10 $\mu$m).}
\end{figure}

Finally, from the results we have obtained, we can gain a deeper understanding of the polarization behavior of the waveguides. We have reported before  that the TM mode is guided, whereas the TE is not \cite{sotillo2016diamond}. The TM mode will have the polarization parallel to the [001] crystallographic direction in the waveguides. So this polarization experiences an increase in the refractive index within the central region between the two barriers (Fig. 4(a)), and the mode is confined. On the other hand, the polarization parallel to [$\bar{1}$10] sees a decrease in the refractive index and thus the TE mode does not meet the conditions to be guided.

In conclusion, we have described the use of polarized Raman spectroscopy to obtain information about the stress distribution and refractive index change in diamond waveguides fabricated with femtosecond laser writing. We have shown that the optical mode is confined in the horizontal direction by the two laser-written lines, whereas in the vertical direction it is confined by the stress-induced refractive index change. The polarization behavior of the waveguides is explained from the stress created in the guiding region. This study provides a useful framework to design the properties of the diamond waveguides by changing the femtosecond laser writing parameters. The method can be extended to analyze stress-induced femtosecond laser written waveguides with different geometries in diamond as well as in other crystals.  

%\subsection{Acknowledgements}
The authors acknowledge support from FP7 DiamondFab CONCERT Japan project, DIAMANTE MIUR-SIR grant and FemtoDiamante Cariplo ERC reinforcement grant. We thank Prof. Guglielmo Lanzani and Dr. Luigino Criante for the use of the FemtoFab facility at CNST - IIT Milano for the laser fabrication experiments. We thank Prof. Roberto Osellame for access to waveguide characterization facilities. We thank Dr. Patrick Salter for helpful scientific discussions.
%\nocite{*}
\bibliography{aipsamp}% Produces the bibliography via BibTeX.

\end{document}